\documentclass[aps,prb,preprint,showpacs,superscriptaddress]{revtex4}
\usepackage{graphicx}
\usepackage{amsmath}
\usepackage{float}
\begin{document}
	
\title{Electric transport as a probe to unveil microscopic aspects of oxygen-depleted YBCO}

\author{C. Acha}
\thanks{corresponding author (acha@df.uba.ar)}
\affiliation{Universidad de Buenos Aires, Facultad de Ciencias
		Exactas y Naturales, Departamento de  F\'{\i}sica, Laboratorio de
		Bajas Temperaturas and IFIBA, UBA-CONICET, Pabell\'on I, Ciudad
		Universitaria, C1428EHA CABA, Argentina}
	
\author{A. Camjayi}
\affiliation{Universidad de Buenos Aires, Facultad de Ciencias
		Exactas y Naturales, Departamento de  F\'{\i}sica, Laboratorio de
		Bajas Temperaturas and IFIBA, UBA-CONICET, Pabell\'on I, Ciudad
		Universitaria, C1428EHA CABA, Argentina}
	
\author{T. Vaimala}
\affiliation{University of Turku, Department of Physics and
		Astronomy, Wihuri Physical Laboratory, FI-20014 Turku, Finland}
	
\author{H. Huhtinen}
\affiliation{University of Turku, Department of Physics and
		Astronomy, Wihuri Physical Laboratory, FI-20014 Turku, Finland}
	
\author{P. Paturi}
\affiliation{University of Turku, Department of Physics and
		Astronomy, Wihuri Physical Laboratory, FI-20014 Turku, Finland}

\date{\today}
\draft

\begin{abstract}

We report on the characterization of Pt-YBa$_2$Cu$_3$O$_{7-\delta}$
 interfaces, focusing on how oxygen vacancies content
($\delta$) affects electrical transport mechanisms. Our study
examines four Pt-YBa$_2$Cu$_3$O$_{7-\delta}$ samples with varying $\delta$ (0.12 $\leq
\delta \leq$ 0.56) using voltage-current measurements across a
temperature range. We successfully model the electrical behavior
using a Poole-Frenkel conduction framework, revealing that
oxygen vacancies create potential wells that trap carriers, directly
influencing conduction. We observe that the energy of these traps
increases as $\delta$ rises, in agreement with a peak previously
detected in optical conductivity measurements. This result supports earlier interpretations, strengthening the proposed connection between oxygen vacancies and the ionization energy associated with impurity bands in oxygen-depleted YBa$_2$Cu$_3$O$_{7-\delta}$.

\end{abstract}


\maketitle

\section{INTRODUCTION}
The YBa$_2$Cu$_3$O$_{7-\delta}$ (YBCO) superconductor has garnered
substantial attention for numerous years, due to its pivotal role in
defining the landscape of high-temperature superconductors,
captivating scientific and technological interest. Its
high-temperature superconducting capabilities have spurred extensive
investigations to unravel the intricate mechanisms governing its
unique properties.~\cite{Tallon23} One of the key enigmas lies in
the role of oxygen stoichiometry, diffusion and ordering,  all
critical parameters influencing both normal-state and
superconducting behaviors, being also factors dictating the diverse
phases observed in YBCO. Its understanding holds implications not
only for YBCO but also for a broad spectrum of oxide materials and
their derived applications. Seminal neutron diffraction studies on
oxygen-depleted YBCO$_{7-\delta}$ samples exhibiting diverse
$\delta$ values~\cite{Jorgensen90}, have elucidated that oxygen
vacancies are mainly located at the O position in the CuO chains
(O(1) sites, following ref.~\cite{Jorgensen90} notation), along the
crystallographic b-axis, establishing a diffusion channel with
minimal impediments.

\begin{figure}[t]
    \vspace{0mm}
    \centerline{\includegraphics[angle=0,scale=0.5]{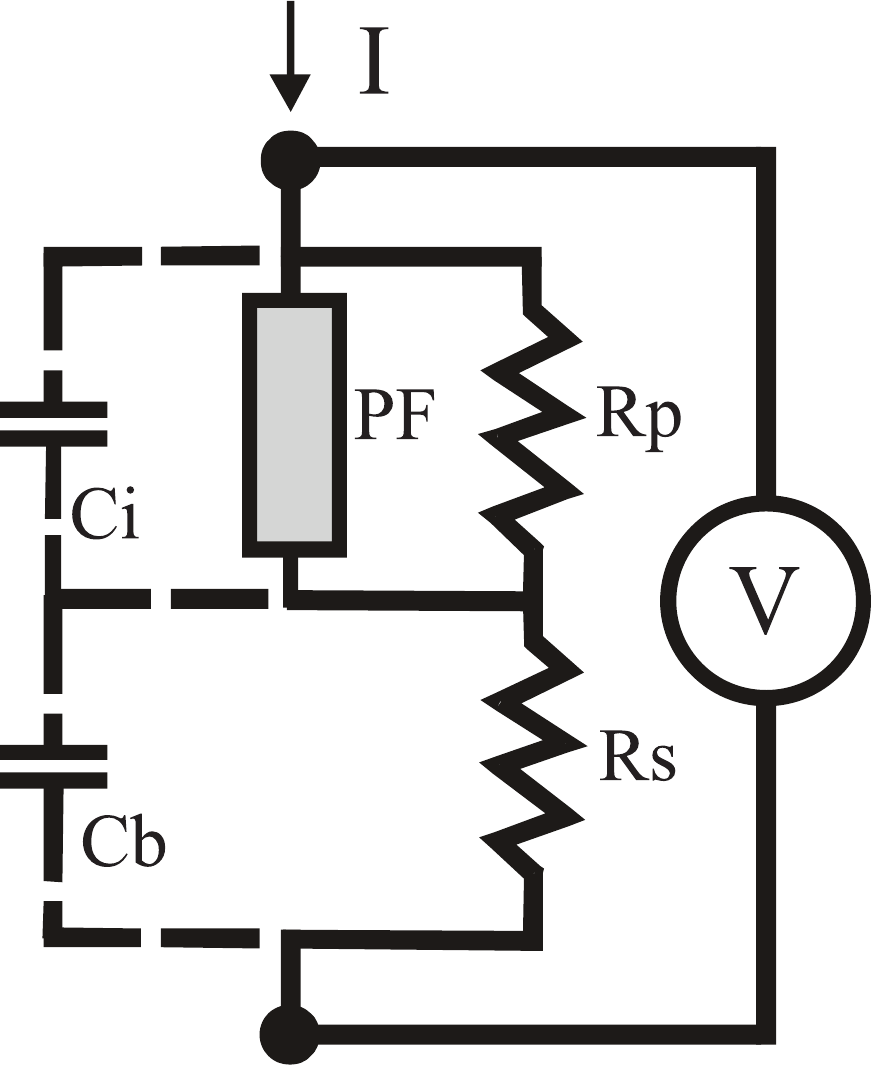}}
    \vspace{0mm}\caption{Equivalent circuit model for the
    metal-transition metal oxide interface. A Poole-Frenkel (PF)
    type conduction is presented here as the particular non-linear
    mechanism. R$_p$ and R$_s$ are the parallel and series
    resistors, respectively (see text).The capacitances C$_i$
    (interfacial) and C$_b$ (bulk) only have a relevant role when
    operating with alternating current at moderate to high
    frequencies. } \vspace{0mm} \label{fig:circuito}
\end{figure}

On a different note, a remarkable progress has already been achieved
in comprehending the physics associated with the mechanisms
governing the write and retention processes in resistive memories,
particularly those linked to oxides where oxygen migration triggers
resistive switching (RS).~\cite{Rozenberg10} This is evident in devices
based on transition metal oxides, like TiO$_2$, Ta$_2$O$_5$,
manganites, cobaltites, and superconducting cuprates, among other
materials. Devices based on metal-YBCO$_{7-\delta}$ may not find
widespread application in memory mainstream technology due to
integration challenges with Si-based electronics and their partial
retentivity~\cite{Schulman12}, attributed to high oxygen diffusivity
in specific crystallographic directions~\cite{Placenik12}.
Nevertheless, their significance lies in unveiling the electrical
transport mechanisms through a metal-complex oxide interface, a
recurring characteristic in many memristive interfaces. Remarkably,
non-recting metal (Au, Pt)-YBCO interfaces have demonstrated interesting bipolar
resistive switching
properties~\cite{Acha09a,Acha09b,Placenik10,Acha11,Schulman13},
coupled with distinctive relaxation
effects~\cite{Schulman11,Placenik12}, electrochemical control of
YBCO's carrier density~\cite{Palau18} and its superconducting performance~\cite{Alcala24}, and peculiar characteristics
associated with inhomogeneous
interfaces~\cite{Schulman12,Schulman15,Waskiewicz15,Truchly16,Waskiewicz18,Tulina18}.
Particularly, the description of its electrical properties has been
achieved through an equivalent circuit  that considers the
Poole-Frenkel (PF) mechanism as responsible for its non-linear
conduction (see Fig.~\ref{fig:circuito}).

It should be noted that neither Au nor Pt cause chemical reactions or ion migration in YBCO. However, they influence the electronic levels through the alignment of the Fermi energies of the metal and the oxide, creating an interfacial region (or depletion layer) with electronic properties distinct from the bulk but still indicative of the oxide's intrinsic characteristics.~\cite{Sze06} The equivalent circuit shown in Fig.~\ref{fig:circuito} also accounts for the distinction of interfacial and bulk regions. The parallel combination of $R_p$ (ohmic) and $R_{PF}$ (the non-linear resistor associated with PF emission) corresponds to the interfacial region. The bulk contribution is captured by the resistor $R_s$. 

Under this description, one might consider that PF traps are solely associated with the interfacial region, with their origin potentially linked, for example, to surface defects. However, it would also be possible that these traps could also exist in the bulk, representing an intrinsic property of YBCO. In the circuital description, they might not be considered because their effect in the bulk can be negligible, as the low bulk resistance ($R_s$) may effectively short-circuit their contribution. In contrast, the interfacial region, with a resistance ($R_p$) comparable to $R_{PF}$, allows for their effects to be more readily observed. This sort of "amplification" effect of the intrinsic bulk oxide properties at the metal-oxide interface was also noted in a previous study of Au-YBCO interfaces.~\cite{Acha09a} In that study, although the contact resistance was 100 times higher than the bulk resistance--which might suggest a YBCO insulating behavior--it actually exhibited metallic characteristics and demonstrated the superconducting resistive transition of YBCO. The resistance variation observed throughout the width of the transition was over an order of magnitude greater than that of bulk YBCO. Additionally, it is worth noting that non-linear IV characteristics were also observed in bulk (4W) measurements of highly deoxygenated YBCO sintered pellets~\cite{Matsushita87} which are consistent with a scenario involving a combination of ohmic and PF-type conduction inherent to YBCO.~\cite{Schulman15}
Within this framework, we aim to elucidate the intrinsic characteristics of electrical transport in oxygen-depleted YBCO by measuring electrical transport across the Pt-YBCO$_{7-\delta}$ interface, where the oxygen content of YBCO has been systematically controlled. This approach aims to correlate the presence of PF traps with oxygen vacancies, enabling us to analyze their specific influence on electrical conduction and providing a unique opportunity to gain deeper insights into the role of these vacancies as carrier traps.

Here, we examine the electrical transport features of 4
Pt-YBCO$_{7-\delta}$ thin film interfaces. By intentionally varying
YBCO's oxygen deficiency during the post-annealing vacuum treatment (0.12 $\leq \delta \leq$ 0.56), our goal is to unravel the $\delta$ dependence of microscopic
parameters linked to PF traps.

\section{EXPERIMENTAL DETAILS}

A series of superconducting YBCO thin films, $\approx 150$\,nm in
thickness, were deposited using pulsed laser deposition (PLD) on
(100) SrTiO$_3$ (STO) substrates. The film growth process utilized a
308\,nm XeCl excimer laser with a pulse duration of 25\,ns and a
repetition rate of 5\,Hz, delivering a laser fluence of
1.6\,J/cm$^2$. The oxygen pressure within the deposition chamber was
maintained at 0.2\,torr, and the substrate temperature during
deposition was set at 750\,$^\circ$C. To optimize the intrinsic film
properties, \textit{in situ} post-annealing treatments were conducted
at a temperature of\,725 $^\circ$C in atmospheric pressure oxygen,
employing heating and cooling rates of 25\,$^\circ$C/min. Further
details regarding the PLD system, growth process, and deposition
parameters can be found elsewhere in comprehensive reports
\cite{Palonen13,Aye21,Khan22}.

The introduction of various oxygen deficiencies and subsequent
variations in the critical temperatures of the samples
\cite{Jorgensen90} was accomplished by annealing the pristine
samples in the pre-chamber of the electron beam evaporator. In this
setup, adjustments could be made to the sample temperature, heat
treatment time, and chamber pressure. The processing parameters for
the final set of samples (labeled S1 to S4) included a constant chamber pressure of
$1.0 \times 10^{-7}$\,mbar and pristine sample annealing
temperatures of 350\,$^\circ$C (S1), 400\,$^\circ$C (S2), 425\,$^\circ$C (S3), and 475\,$^\circ$C (S4). The
heating and cooling rates for these treatments were set at
10\,$^\circ$C/min, and the duration of each treatment at the target
temperature was 60\,s.

\begin{figure}[htbp]
    \vspace{0mm}
    \centerline{\includegraphics[angle=0,scale=0.6]{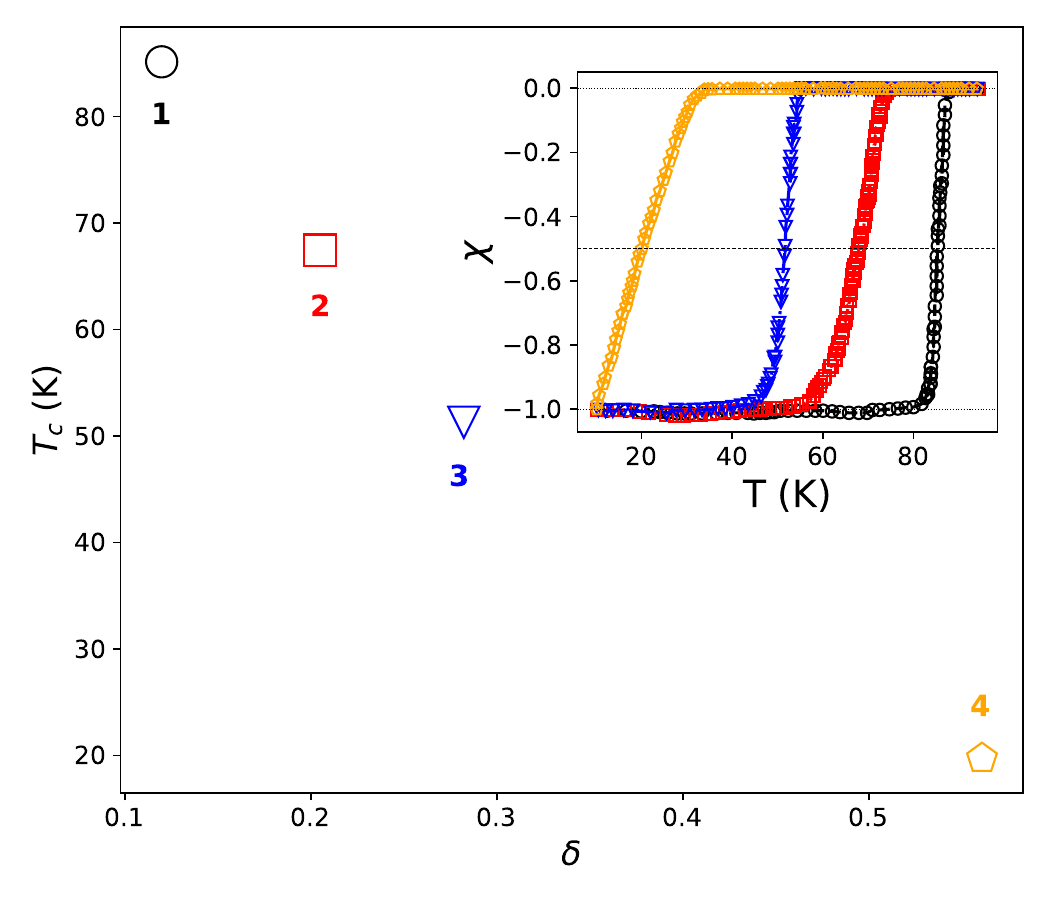}}
    \vspace{0mm}\caption{ Critical superconducting temperature
    ($T_c$, determined by the criterion of the 50\% of the magnetic susceptibility $\chi$
    transition, shown in the inset) as a function of the oxygen deficiency ($\delta$),
    identified through the ratio between the 2-$\theta$ (005) and
    (004) peaks from characterizations using XRD in each film. Next to each point,
    the numbers associated with each of the samples are indicated.}
    \vspace{0mm} \label{fig:Tcvsdelta}
\end{figure}

The ultimate oxygen deficiencies were determined by calculating the
ratios of the x-ray diffraction (XRD) (005) and (004) peak
intensities, denoted as $I(005)/I(004)$, as extensively detailed in
ref. \cite{Ye93} (see details in the Supplementary Material
section). The superconducting critical temperatures ($T_{\mathrm
c}$) were derived from the temperature-dependent susceptibilities,
as depicted in the inset of Fig.~\ref{fig:Tcvsdelta}. Here,
$T_{\mathrm c}$ was determined using the criterion of 50\% of
$\chi$, aligning closely with the first-order derivative of the
susceptibility curve. The observed experimental relationship between
critical temperature and oxygen deficiency is presented in
Fig.~\ref{fig:Tcvsdelta}, demonstrating a consistent alignment with
the previously described dependence observed in oxygen-deficient
YBCO powder samples \cite{Jorgensen90}.

\begin{figure}[t]
\vspace{0mm}
\centerline{\includegraphics[width=0.8\linewidth]{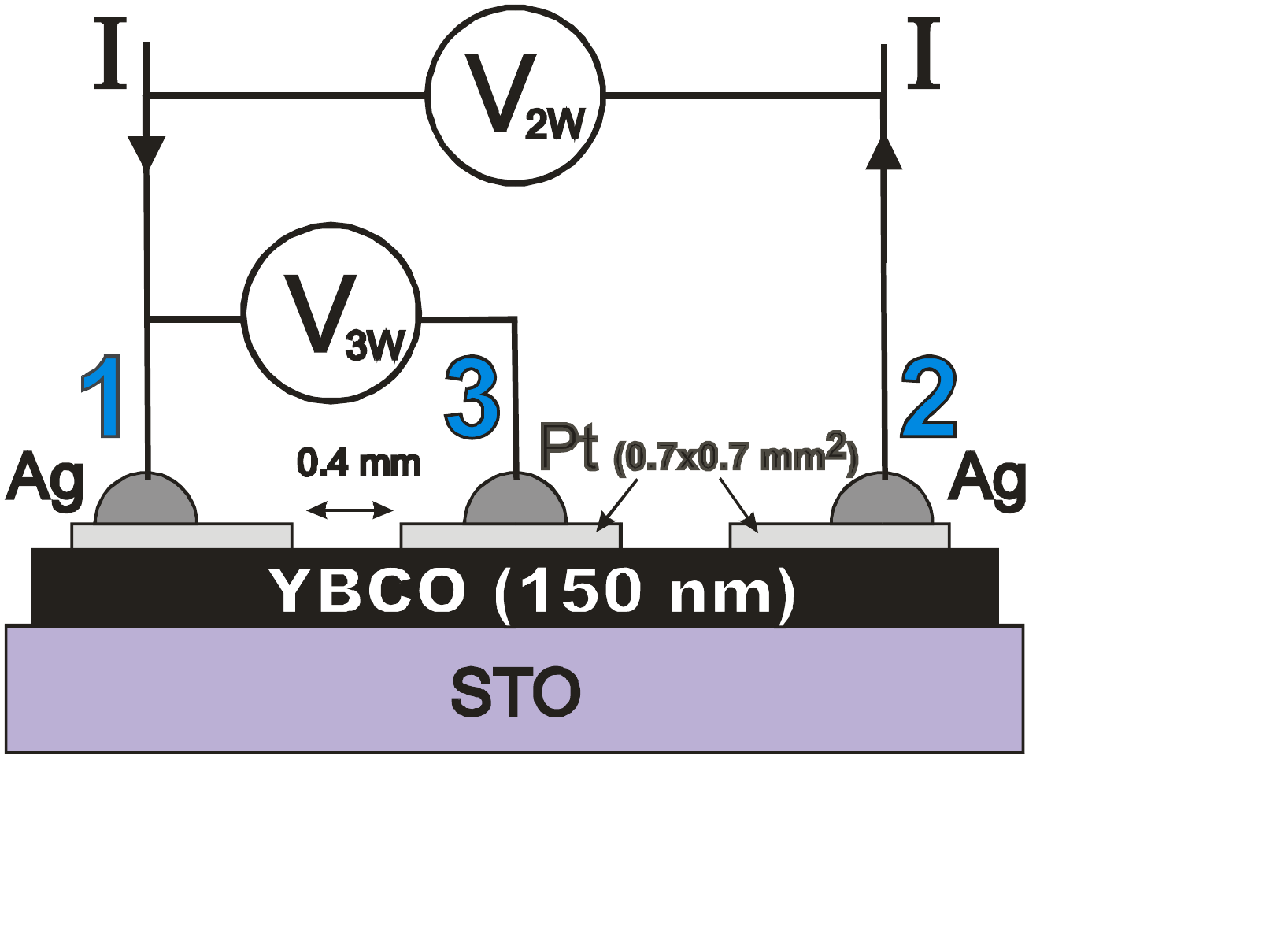}}
\vspace{-10mm} \caption{ Schematic of the studied
Pt-YBCO$_{7-\delta}$ devices, showing the implemented
    electrical contacts distribution and the employed 2W and 3W setups. }
\vspace{0mm} \label{fig:contactos}
\end{figure}

In order to electrically characterize the Pt-YBCO$_{7-\delta}$
interfaces, several electrodes were deposited on top of each
YBCO$_{7-\delta}$ thin film. This process involved employing
photolithography and depositing high-purity Pt by sputtering
technique. The resulting Pt-sputtered electrodes possess a narrow
thickness of 30 nm, cover an area measuring 0.7 x 0.7 mm$^2$, and
maintain a consistent separation of 0.4 mm. Subsequent to this
fabrication, Pt leads were affixed using silver paint, ensuring a
meticulous connection without direct contact with the YBCO sample
surface (see Fig.~\ref{fig:contactos}).

To characterize the current-voltage (IV) response of the Pt/YBCO
interfaces at different temperatures, 1 ms voltage pulses were
applied between two Pt electrodes, labeled as "1" and "2"
respectively, systematically cycling their amplitude using an
Agilent B2902B SMU. Pulses were applied instead of a DC current to limit local Joule dissipation and thus minimize the possibility of RS occurrence, which could alter the oxygen vacancy distribution, primarily in the interfacial region where most of the applied voltage drops. The cycling involved varying the voltage between
two extreme values, starting from the negative value and reaching
the positive one in each cycle. The contact design was configured to
facilitate both 2-Wire (2W) and 3-Wire (3W) measurements by
introducing an additional electrode ("3") between the aforementioned
ones, allowing for a comprehensive characterization of each Pt/YBCO
interface. Simultaneously, the instrument quantified the circulating
current (I) throughout each pulse and measured the potential
difference across the electrodes $V_{1-2}$ (2W) and $V_{1-3}$ (3W).
Introducing a $V_{bias}=0.1V$ pulse between each main pulse allowed
us to measure the remnant resistance, denoted as $R_{rem}$. Each IV
measurement was conducted with temperature stabilization better than
250 mK within the overall range of 80 K to 300 K, with steps ranging
from 5 K to 10 K. It is worth noting that the lower limit of the
measured temperature for each film was restricted to guarantee the
execution of low-noise measurements, when the interface resistance
becomes too large. In addition, the voltage range explored for each
device was tailored to prevent a RS at any temperature. To achieve
this, a 'safe' voltage range was initially used to obtain IV
characteristics for all studied temperatures. The range was gradually extended, repeating the study for all temperatures at each step. Once an RS occurred, the experiment was terminated at that point and for all further temperatures.
Thus, the study was restricted to IV curves that did not exhibit RS
throughout the entire temperature range.

\section{RESULTS AND DISCUSSION}

\begin{figure}[htbp]
    \vspace{0mm}
    \centerline{\includegraphics[angle=0,scale=0.4]{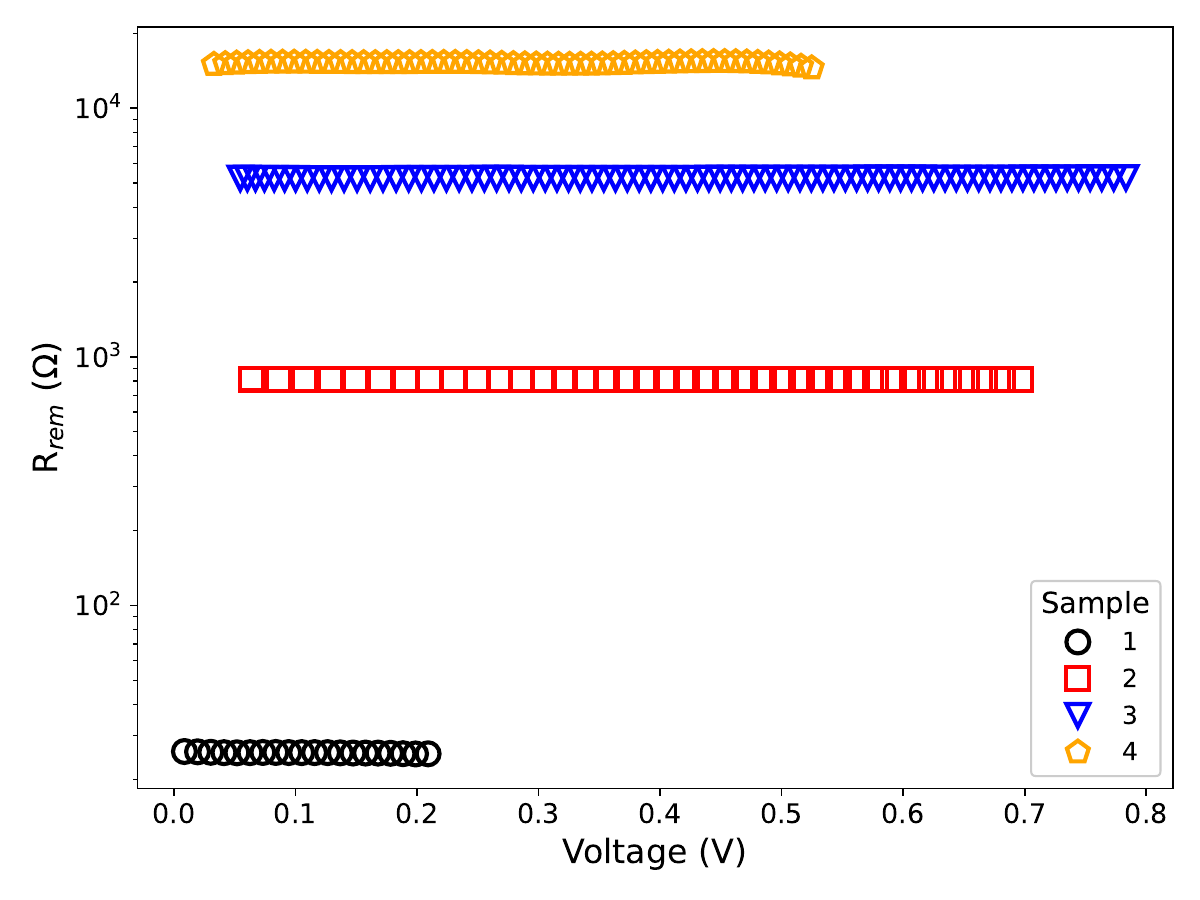}}
    \vspace{0mm}\caption{Remanent resistance ($R_{rem}$) as a function
        of write voltage ($V$). Since no dependency is observed, we are
        confident that the resistive state of the interfaces was not altered
        during the study of the IV characteristics.} \vspace{0mm}
    \label{fig:Rremxx}
\end{figure}

\begin{figure}[htbp]
\vspace{0mm}
\centerline{\includegraphics[angle=0,scale=0.4]{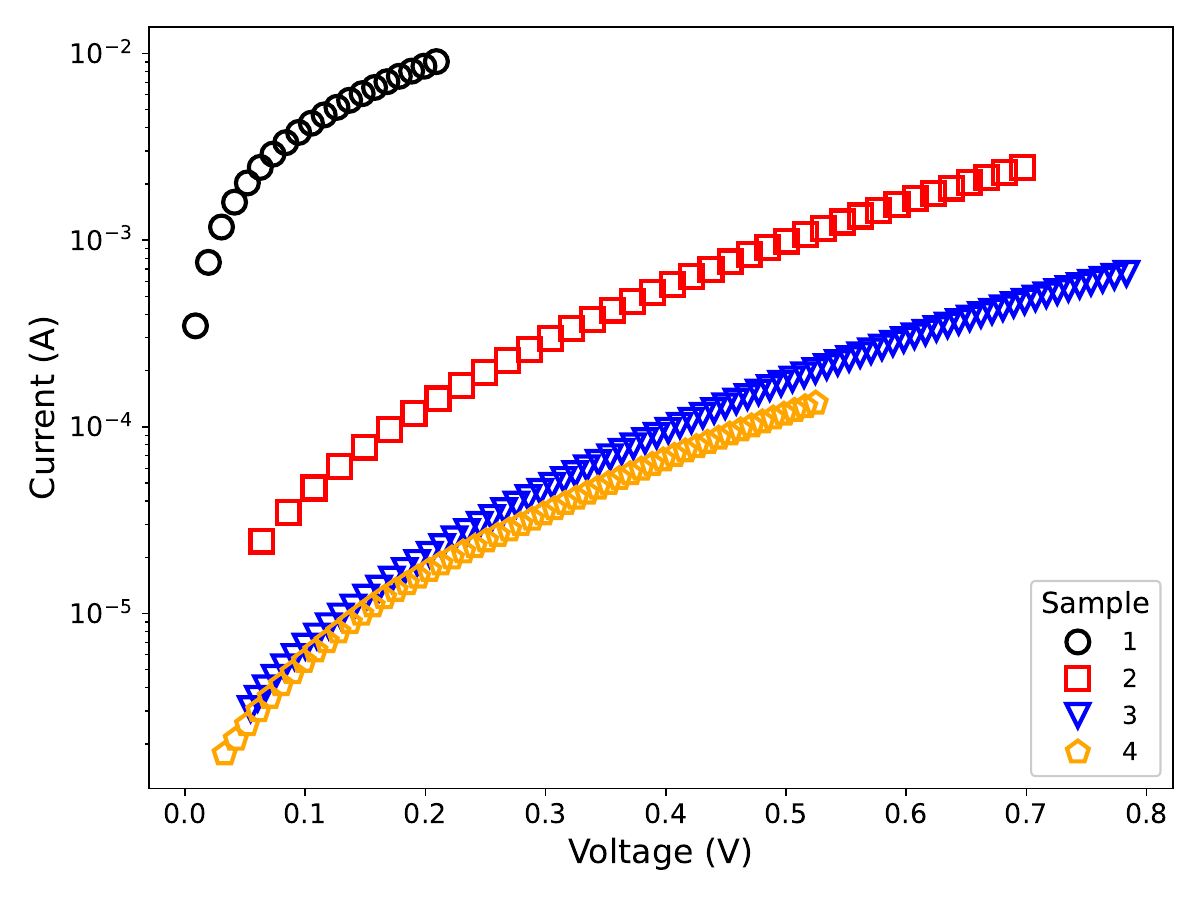}}
\vspace{0mm}\caption{Comparison of the IV characteristics of samples
1-4  at 290 K, within a limited range of voltages and currents to
prevent resistive changes that could alter the original oxygen
content in the vicinity of the contacts.} \vspace{0mm}
\label{fig:IVxx290}
\end{figure}

In Figs.~\ref{fig:Rremxx} and ~\ref{fig:IVxx290}, the constant
preservation of remanent resistance $R_{rem}$ and the IV
characteristics for all studied Pt-YBCO$_{7-\delta}$ interfaces at
290 K, within the utilized write voltage range are displayed,
respectively.

The Pt-YBCO$_{7-\delta}$ devices exhibit distinct conduction levels,
with nearly 3 orders of magnitude difference between the one based
on the most oxygenated YBCO (S1) and the most deoxygenated (S4). The
``safe'' voltage and temperature range---\textit{i.e.}, without
causing a RS that could alter YBCO's oxygen distribution---was
highly variable across the samples, generally featuring a
non-monotonic voltage scale, which was significantly reduced for the
most conductive one (S1). Concerning the temperature range,
the minimum value progressively increased as the samples became less
conductive, severely limiting the range for S4.

\begin{figure}[htbp]
    \vspace{0mm}
    \centerline{\includegraphics[angle=0,scale=0.4]{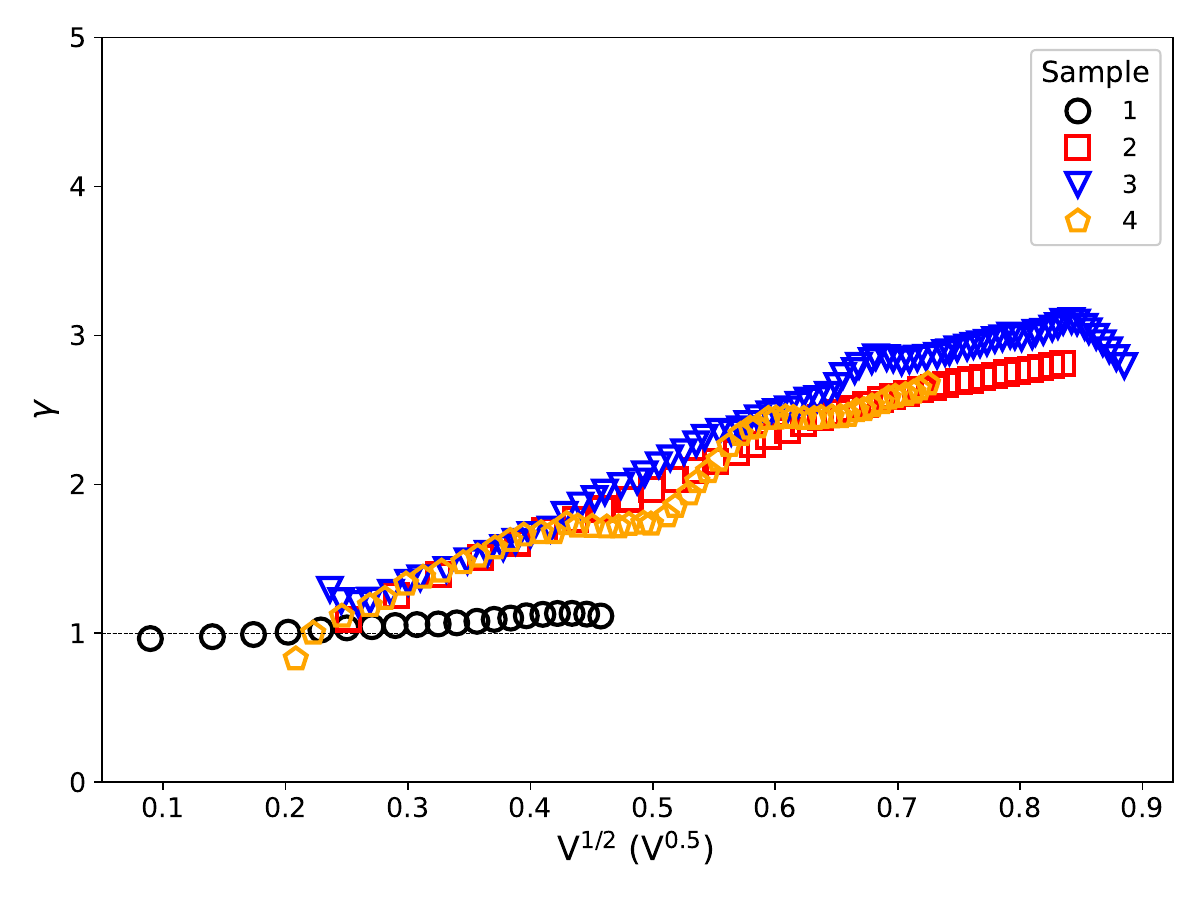}}
    \vspace{0mm}\caption{Comparison of the parameter $\gamma$ obtained
        for S1 to S4 at 290 K. A complex behavior is
        observed indicating the simultaneous existence of different
        conduction mechanisms, marked by both the varying degree of
        non-linearity ($\gamma > 1$) and the existence of ohmic processes,
        both in parallel ($\gamma \simeq 1$ at low voltages) and in series
        (maximum in $\gamma$ for the highest voltages).} \vspace{0mm}
    \label{fig:gammaxx}
\end{figure}

It can be observed that the functional dependence of the IV curves
varies for each interface. To highlight these differences more
effectively, the graphical representation of the power exponent
$\gamma=dLn(I)/dLn(V)$ plotted against $V^{1/2}$, shown in
Fig.~\ref{fig:gammaxx}, can be employed.~\cite{Acha17}This method
has proven highly valuable in identifying different transport
mechanisms when multiple mechanisms contribute to electric
transport.~\cite{Acha16,Acevedo17,Ghenzi19} Distinct degrees of
non-linearity are visible, consistent with already mentioned
existence of diverse conduction mechanisms, typically observed in
metal-YBCO junctions, whose circuit representation, including a PF
non-linear element, is depicted in Fig.~\ref{fig:circuito}.

\begin{figure*}[htbp]
    \vspace{0mm}
    \centerline{\includegraphics[angle=0,scale=0.45]{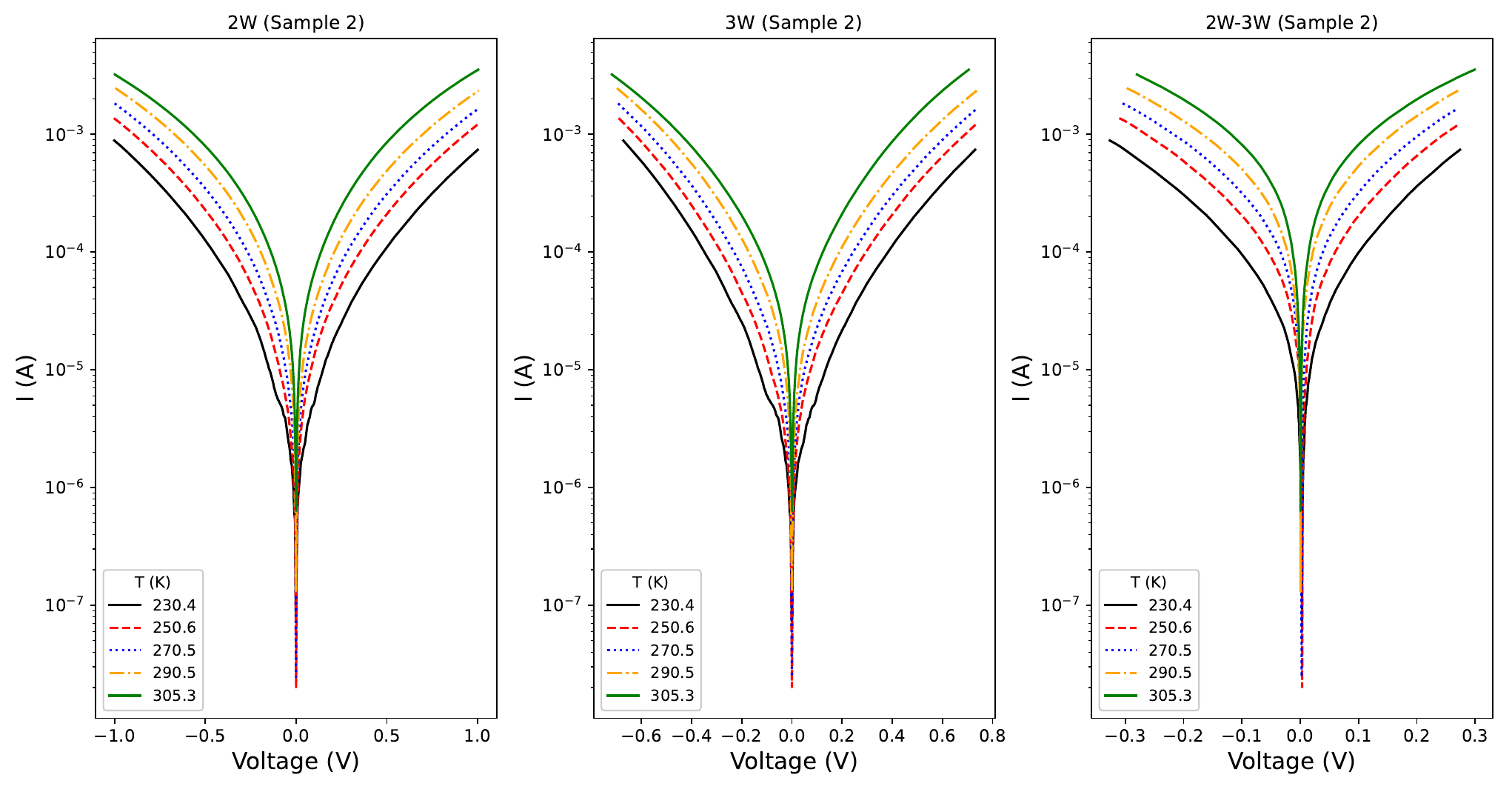}}
    \vspace{0mm}\caption{IV characteristics as a function of
        temperature measured using either 2W (2 Pt-YBCO interfaces) or
        3W (1$^{st}$ Pt-YBCO interface) configurations for S2. The
        2W-3W is the mathematical difference of these measurements,
        indicating the IV dependence of the 2$^{nd}$ interface, not
        measured within the 3W configuration. Not appreciable
        rectification effects are observed, besides a small voltage
        asymmetry between the 3W and the 2W-3W measurements, which may
        be associated with the overall resistance of the external
        Cu-Ag-Pt pads.} \vspace{0mm} \label{fig:IV2}
\end{figure*}

Fig.~\ref{fig:IV2} illustrates the temperature sensitivity of the IV
characteristics for S2, measured under both the 2W and 3W
configurations. Similar results were obtained for the other samples.
The electrical conductivity is observed to increase with rising
temperature. A subtle rectification is noted but will be disregarded
in the modeling process to maintain a straightforward and convenient
description. Quantitative analysis will concentrate on the outcomes
from the 3W configuration for all samples, ensuring a measurement
solely of the contribution from one of the Pt-YBCO$_{7-\delta}$
interfaces rather than both.

Based on prior findings which indicated that the equivalent circuit
shown in Fig.~\ref{fig:circuito} is the optimal representation of
the Pt-YBCO$_{7-\delta}$ interface~\cite{Schulman15, Lanosa20}, we
can describe its DC electrical behavior using the following
equations~\cite{Hill71,Vollmann74}:

\begin{equation}
\label{eq:PF} I_{PF}(V_{PF}) = A~V_{PF}~exp\left[ C
\sqrt{V_{PF}}\right],
\end{equation}

\begin{equation}
\label{eq:A} A =  A_{PF}~\exp \left[ \frac{ - {\phi}_T}{T} \right],
\end{equation}

\begin{equation}
\label{eq:Apf} A_{PF} = \frac{S}{d}~q~\mu~(N_CN_D)^{1/2},
\end{equation}

\begin{equation}
\label{eq:C} C = \frac{q^{3/2}}{k_B T (\pi \epsilon^{'}\epsilon_0
d)^{1/2}},
\end{equation}

\noindent where $I_{PF}$ and $V_{PF}$ are the current and the
voltage across the PF element, respectively. $S$ is the cross
section area of the conducting path, $q$ the electron's charge,
$\mu$ the electronic drift mobility, $N_C$ and $N_D$ the density of
states in the conduction band and the donor density, respectively.
$\phi_T$ is the trap energy level (in K), $k_B$ the Boltzmann
constant, $\epsilon^{'}$ the real part of the relative dielectric
constant of the oxide, $\epsilon_0$ the permittivity of vacuum and
$d$ the distance associated with the voltage drop $V_{PF}$. Notice
that $d$ may not necessarily be equivalent to the distance between
the voltage contacts. As $ V_{PF} = V - I R_s $ and $I = I_{PF} +
I_{Rp}$, where $R_s$ and $R_p$ are the series and the parallel
resistances, respectively, then:

\begin{equation}
\label{eq:I-V} I = A(V - I~R_s)~exp\left[C~\sqrt{V - I~R_s} \right]
+ \frac{V - I~R_s}{R_p}.
\end{equation}

Equation~\ref{eq:I-V} is an implicit equation requiring numerical
solution for fitting the experimental data. For this purpose, we
arbitrarilly choose the IV data with positive $V$ values and we
applied an optimization technique that involved a generalized
reduced gradient method for the $A$ and $C$ parameters, along with a
'brute force' variation of the $R_p$ and $R_s$ parameters. Testing
different initial values for these parameters allowed us to identify
the solution with the lowest residuals. We verified that the
obtained parameters for this solution (within a 10\% range)
consistently minimized residuals when fitting the data from the
negative $V$ quadrant.

\begin{figure*}[htbp]
\vspace{0mm}
\centerline{\includegraphics[width=16cm,height=8cm]{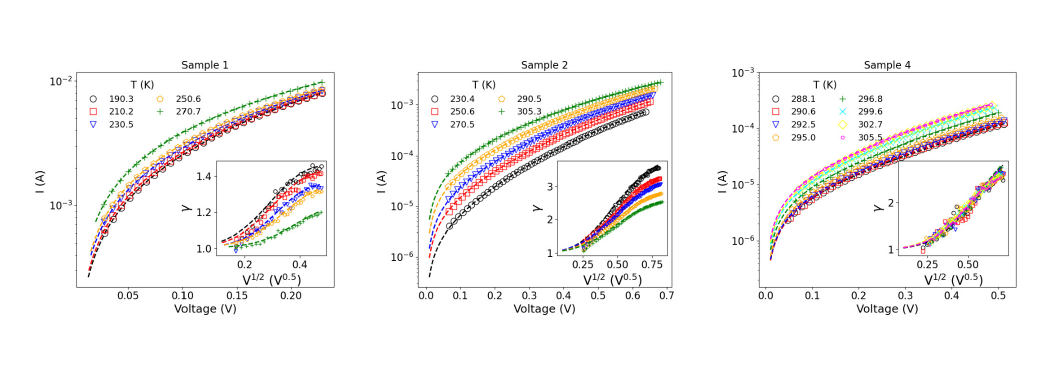}}
\vspace{-10mm}\caption{IV characteristics at different temperatures
for S1, S2 and S4. The dashed lines are fits using the electric
circuit model of Fig.~\ref{fig:circuito} associated with
Eq.~\ref{eq:I-V}. The inset shows the dependence of $\gamma$ on
V$^{1/2}$ at different temperatures for the experimental data, where
the dashed lines represent the obtained $\gamma$ for the fitting IV
curves, also showing an excellent agreement.  Similar results were
obtained for S3 (see Supplementary Material).} \vspace{0mm}
\label{fig:IV_gamma_ajuste}
\end{figure*}

\begin{figure}[htbp]
    \vspace{0mm}
    \centerline{\includegraphics[angle=0,scale=0.7]{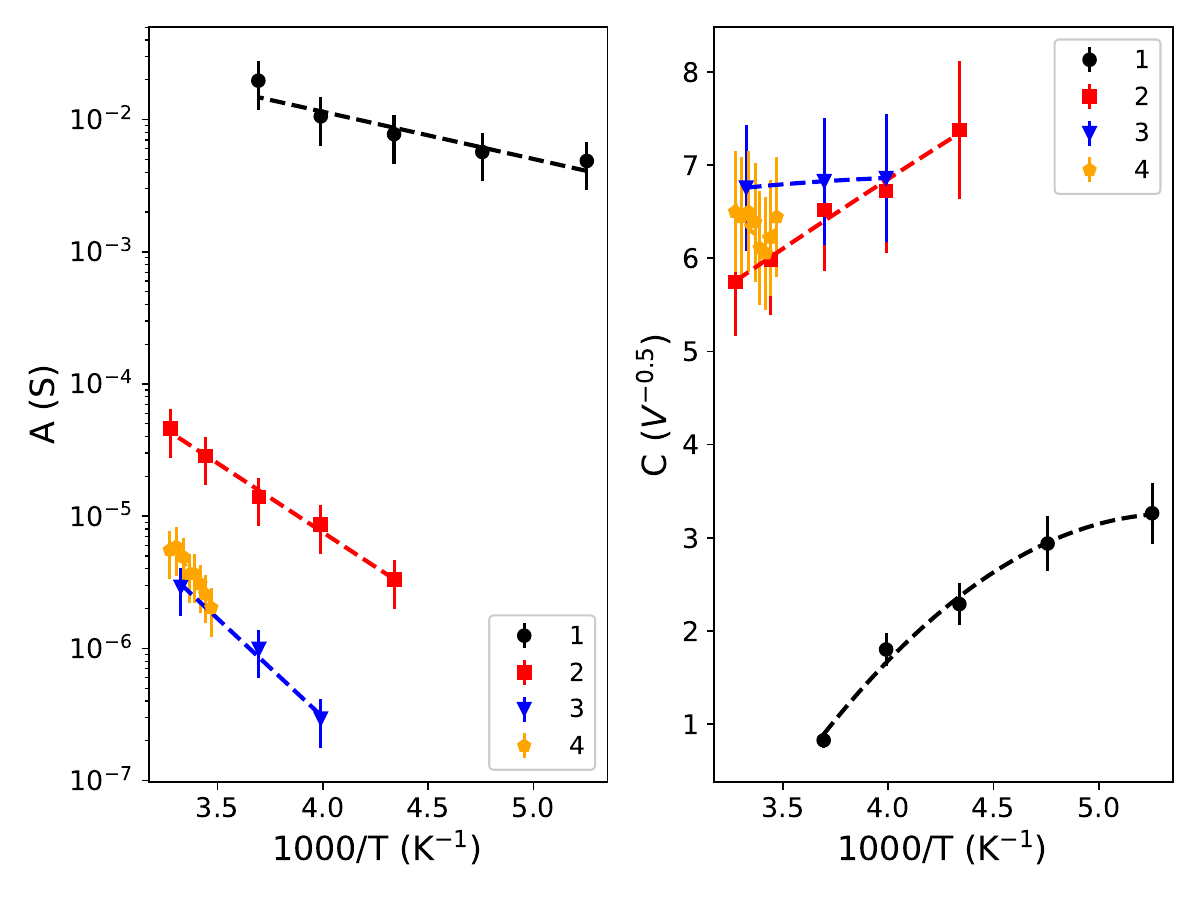}}
    \vspace{0mm}\caption{ The obtained $A$ and $C$ Poole-Frenkel
    parameters after fitting the IV characteristics as a function of
    temperature for all the samples. Dashed lines for the $A$ parameter are fits using Eq.~\ref{eq:A}
    while they are guides to the eye for $C$ parameter. } \vspace{0mm} \label{fig:AyC}
\end{figure}

As can be observed in Fig.~\ref{fig:IV_gamma_ajuste}, the
experimental IV characteristics for S1, S2 and S4 are very well
reproduced by the theoretical representation of the proposed circuit
model (Fig.~\ref{fig:circuito} and Eq.~\ref{eq:I-V}), respectively.
The quality of the fits can also be appreciated in the
representation of the $\gamma$ curves, shown as an inset in
Fig.~\ref{fig:IV_gamma_ajuste}). A small deviation can be observed
at low currents and voltages, likely attributed to the presence of
minimal thermoelectric voltages at the interfaces as well as well as
the minor rectifying effects already mentioned. Similar results were
obtained for S3 (see Supplementary Material).

\begin{figure}[h]
\vspace{-3mm}
\centerline{\includegraphics[angle=0,scale=0.7]{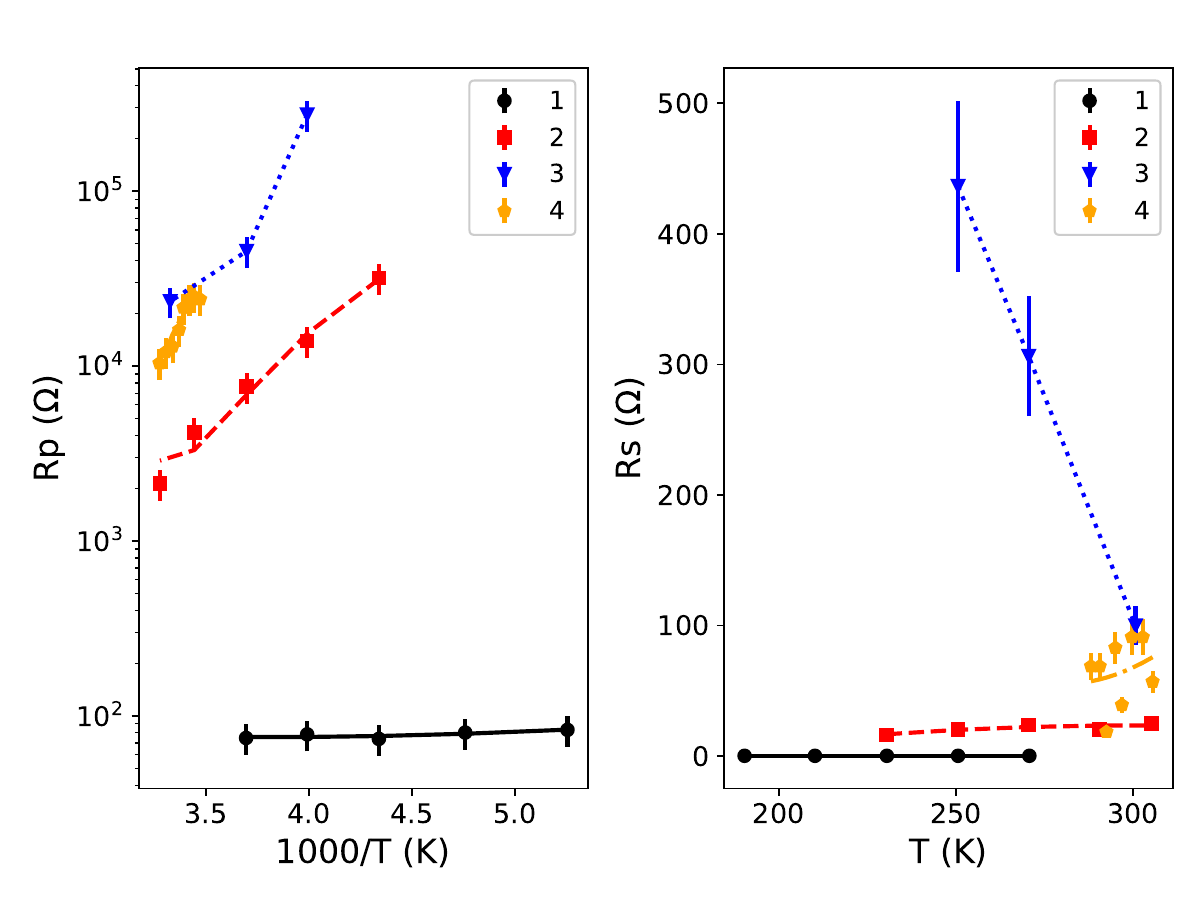}}
\vspace{0mm}\caption{The obtained parallel ($R_p$) and series
resistance ($R_s$) after fitting the IV characteristics as a
function of temperature for all the samples. Lines are guides to the
eye.} \vspace{0mm} \label{fig:RpyRs}
\end{figure}

The parameters of the best fits can be observed in
Figs.\ref{fig:AyC} and Fig.\ref{fig:RpyRs}. For parameter $A$, the
expected dependence according to Eq.\ref{eq:A} is obtained,
confirming that the term $A_{PF}$ remains essentially temperature-independent
within the studied range and varies with the oxygen content of YBCO.
Regarding parameter $C$, the
only sample showing consistent behavior with Eq.\ref{eq:C} is S2.
This is strongly linked to the temperature dependence that the
$\epsilon'$ associated with this PF-type conduction zone may
exhibit. This indicates that S2 presents a temperature-independent
$\epsilon'$, while this is not the case for the other samples.

The behavior of the parallel resistance, $R_p$ (see
Fig.~\ref{fig:RpyRs}), is consistent with previous
results~\cite{Schulman15,Lanosa20}, where we observed that it can be
considered as an ohmic leakage channel. It corresponds to a
semiconductor-type conduction, more specifically that associated
with disorder (variable range hopping). In general, it is observed
that with greater deoxygenation, both the resistance value and its
slope (vs. 1/T) increase. On the other hand, the series resistance,
$R_s$, is metallic-like for the more oxygenated samples (S1 and S2)
and semiconducting for S3. The behavior of S4 is not quite clear,
probably as a result of the high noise level (low currents) and the
constrained temperature range studied.

\begin{figure}[h]
\vspace{0mm}
\centerline{\includegraphics[angle=0,scale=0.7]{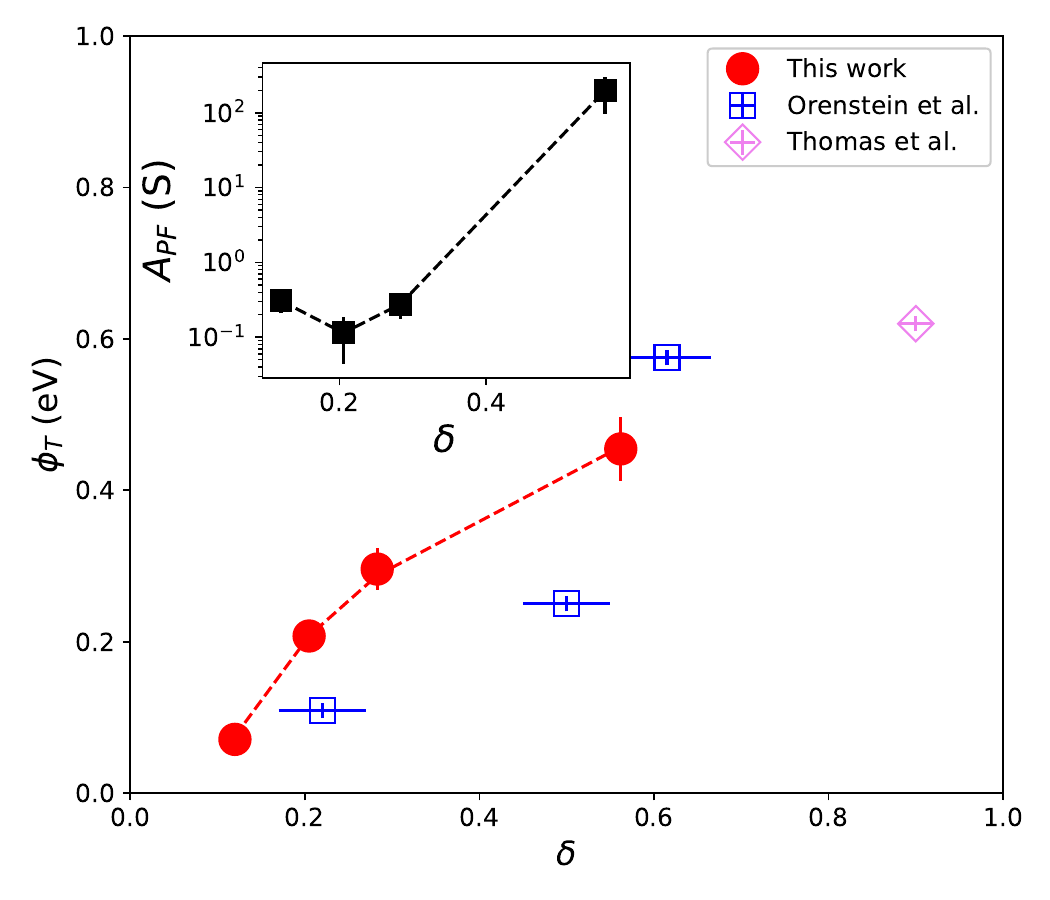}}
\vspace{0mm}\caption{Oxygen deficiency ($\delta$) dependence of the
trap energy [$\phi_T(\delta)$] (red solid circles). $E_I(\delta)$
extracted from optical conductivity measurements from Orenstein et.
al. (ref.~\cite{Orenstein90}) and Thomas et al. (ref.~\cite{Thomas92})  (blue squares and violet diamond, respectively). The inset shows the
$\delta$ dependence of the Poole-Frenkel pre-factor
[$A_{PF}(\delta)$]. Dashed lines are guides to the eye.}
\vspace{0mm} \label{fig:A y Phi}
\end{figure}

Finally, the dependence of parameters $A_{PF}$ and $\phi_T$ on
$\delta$, associated with Eq.\ref{eq:A}, is depicted in
Fig.~\ref{fig:A y Phi}. It is observed that $A_{PF}$ exhibits a
non-monotonic behavior and a sharp rise for $\delta$ = 0.56, whereas
$\phi_T$ shows a logarithmic-type increase.
We understand that the
complex response of the parameter $A_{PF}$ is associated with its
multiple dependencies, as it relates both to electronic parameters
(donor density, density of states in the conduction band, carrier
mobility) and geometric parameters (the conduction surface and the
thickness of the interfacial region, see Eq.~\ref{eq:Apf}). This competition among parameters complicates the inference of the expected behavior with varying $\delta$. However, we understand that the significant increase observed at the highest $\delta$ likely indicates a substantial shift in detrapped carrier mobility or changes in geometric factors, such as an expanded effective conduction area or a reduced interfacial zone.
The dependence of $\phi_T$ on $\delta$ confirms that the trap
potential well affecting PF conduction deepens as the concentration
of oxygen vacancies in YBCO increases. This observation validates
the role of oxygen vacancies in generating traps for charge carriers
traveling through YBCO. \par \noindent The established empirical
$\phi_T(\delta)$ relation can be used to infer the oxygen content
from electrical transport measurements.  In previous
work~\cite{Schulman15}, it was shown that $\phi_T$ could be altered
by applying either higher amplitude pulses or a greater number of
pulses of the same amplitude. After applying pulses that caused a
200\% increase in the initial resistance of an Au-YBCO device,
$\phi_T$ increased from approximately 0.06 eV to 0.11 eV. From
Fig.~\ref{fig:A y Phi}  we can estimate that  this represents a
local change in delta from $\simeq$~0.11 to 0.13. These values are in
agreement with the studies conducted in that work on the temperature
dependence of the resistance of those interfaces, as no signs of electronic
localization (VRH) were observed, which would have been expected for higher
values of $\delta$. This consistency highlights the viability of employing
electrical transport as a tool to assess
the oxygen content in the interfacial zone of the metal-YBCO
junction.\par

To better understand the implications of the observed
$\phi_T(\delta)$ dependence, we might examine studies involving
optical conductivity, which has played a key role in disentangling
the complexities of electronic band structures and basic
excitations. Indeed, by conducting optical reflectivity,
ellipsometry, and Raman scattering studies on YBCO$_{6+x}$ (0 $\leq
x \leq$ 1) samples, Cooper et al.~\cite{Cooper93} compared findings
from twinned YBCO single crystals to those single-domain, as well as
from BSCCO crystals (which do not contain CuO chains). This allowed
them to distinguish the contributions to optical conductivity
arising from both CuO$_2$ plane transitions and CuO chains. One of
their key observations was that in samples where $x < 0.8$ (i.e.
$\delta > 0.2$), a mid-infrared peak is detected in the electrical
conductivity, shifting towards lower energies as oxygen vacancies
are reduced. This absorption, which is solely attributed to the CuO
chains where the oxygen vacancies are located, leads to a deviation
from the observed low-frequency dependence ($\sigma \sim
\omega^{-1}$), potentially indicating the presence of an electronic
band associated with the vacancies. In other words, it seems
reasonable to associate this region of increased absorption with the
set of energies required to ionize an impurity charge
(E$_I$)~\cite{Mott90}, that is, to extract a charge carrier from the
potential well associated with the oxygen vacancy and promote it to
the conduction band or to similar
energy bands. \\

Thomas et al. (ref.~\cite{Thomas92}) analyzed the optical
conductivity of several lightly doped cuprates (including YBCO$_x$
with x$\simeq 6.1$), revealing the presence of two peaks for
energies below 1 eV. The plausibility of associating the higher
energy peak with E$_I$ was discussed. Similarly, we have extracted
from ref.~\cite{Orenstein90} the energies of the maxima in
$\sigma(\omega)$ for YBCO$_{7-\delta}$ samples with 30 K $\leq T_c
\leq$ 80 K. We were able to correlate T$_c$ with $\delta$ based on
studies that defined T$_c$ similarly (onset of the magnetization
curve at low field).~\cite{Jorgensen90,Cava90}. A detailed description of the procedure we applied can be found in the Supplementary Information section. These derived
$E_I(\delta)$ points have been added to Fig.~\ref{fig:A y Phi},
showing a strong correlation with $\phi_T(\delta)$. This strengthens
the initial interpretation regarding the origin of this optical
conductivity peak as an ionization energy.

\section{CONCLUSIONS}
We have measured the electrical transport across Pt-YBa$2$Cu$3$O${7-\delta}$ (0.12 $\leq \delta \leq$ 0.56) interfaces at various temperatures. Results were analyzed by considering a circuital model, previously demonstrated to be effective in describing transport in optimally doped metal-YBCO junctions. The model features two interfacial component --a PF emission-based element in parallel with a resistor-- in series with a bulk resistance (R$_s$) considerably smaller than R$_p$, highlighting the dominance of interfacial effects in transport properties. Since Pt electrodes are chemically inert with respect to YBCO, the observed PF emission can be attributed to intrinsic YBCO properties linked to oxygen vacancies. Importantly, the microscopic parameters governing PF emission--notably the trap energy--demonstrated a systematic dependence on oxygen content. As $\delta$ increased, so did the energy of the potential wells, a trend that correlates strongly with a peak identified in optical conductivity studies. This reinforces the proposed relationship between oxygen vacancies and the ionization energy of charges localized within impurity bands, offering deeper insights into the microscopic transport mechanisms in oxygen-depleted YBCO.

\section{ACKNOWLEDGEMENTS}
We acknowledge very useful discussions with Dr. V. Ferrari and Dr.
J. Dilson, as well as financial support by CONICET Grant PIP
11220200101300CO and UBACyT 20020220200062BA (2023-2025). Jenny and
Antti Wihuri Foundation is also acknowledged for financial support.

\pagebreak
\section{\large Supplementary Information:}

\subsection{Oxygen deficiency determination}

\begin{figure}[htbp]
	\vspace{0mm}
	\centerline{\includegraphics[angle=0,scale=0.8]{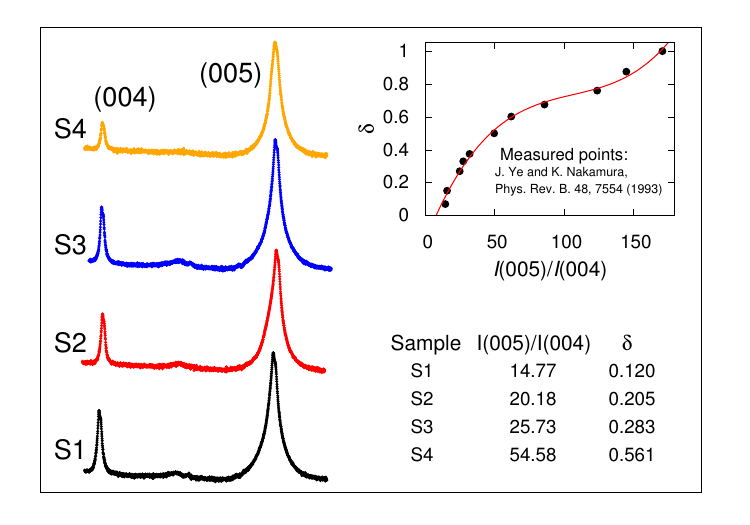}}
	\vspace{0mm}\caption{XRD of S1-S2-S3-S4 focused on the 2-$\theta$ (005) and
		(004) peaks (left side). The $\delta$ dependence on the
		I(005)/I(004)peak intensity ratio, from ref.~\cite{Ye93} (upperside
		right) and the table with the obtained values (downside right).}
	\vspace{0mm} \label{fig:Xray}
\end{figure}

\begin{enumerate}
	\item The intensity ratio between the XRD peaks (005) and (004),
	I(005)/I(004), is calculated from the peak areas shown on the
	left-hand side of Fig.~\ref{fig:Xray}. This intensity ratio is also
	presented in the table at the bottom right.
	\item The relationship between the oxygen deficiency, $\delta$, and the
	intensity ratio I(005)/I(004) is derived from Ye's and Nakamura's
	paper~\cite{Ye93}, as shown in the figure inset. The data points in
	the inset are taken from the paper, and we have fitted a third-order
	polynomial function to describe this relationship.
	\item By inputting the measured intensity ratios into the polynomial
	function, the corresponding oxygen deficiencies (delta) can be
	calculated. These values are also shown in the table at the bottom
	right.
	\item The measured critical temperature values can then be plotted as a
	function of the oxygen deficiency, as presented in
	Fig. 2 (main text).
\end{enumerate}

\subsection{\large IV characteristics of S3}

\begin{figure}[htbp]
	\vspace{0mm}
	\centerline{\includegraphics[angle=0,scale=0.55]{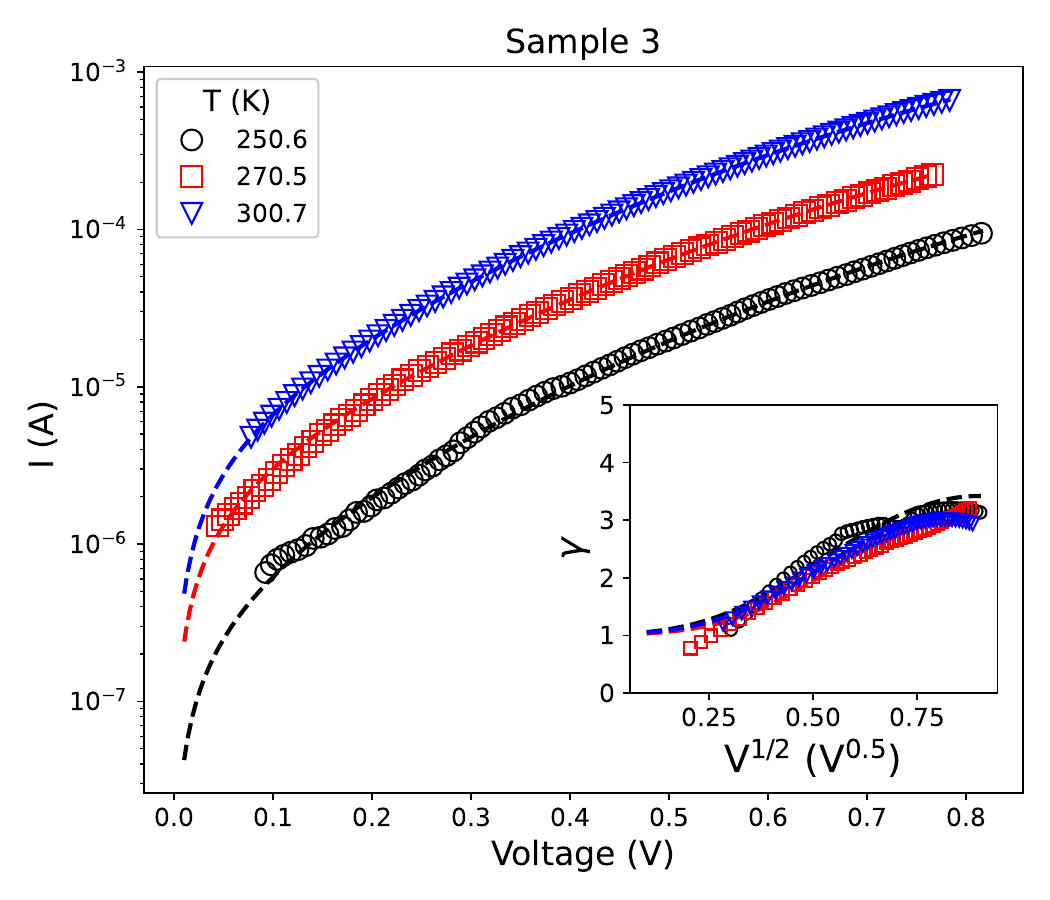}}
	\vspace{0mm}\caption{IV characteristics at different temperatures
		for S3. The dashed lines are fits using the electric circuit model
		of Fig.~1 associated with Eq.~5. The inset shows the dependence of
		$\gamma$ on V$^{1/2}$ at different temperatures for the experimental
		data, where the dashed lines represent the obtained $\gamma$ for the
		fitting IV curves, also showing a good agreement.}
	\vspace{0mm} \label{fig:3_IV}
\end{figure}

The IV characteristics as well as the $\gamma$ parameter for S3, not
included in the main text for simplicity, are shown in
Fig.~\ref{fig:3_IV}.

\subsection{\large $E_I(\delta)$ extracted from Orenstein et al.~\cite{Orenstein90} }

Thomas et al.~\cite{Thomas92} studied the optical conductivity of several lightly doped semiconductors derived from cuprates, among which YBa$_2$Cu$_3$O$_{6.1}$ was included. Their analysis revealed that for energies lower than 1.25 eV, all spectra shared common characteristics, most prominently two peaks attributed to the energies labeled $E_J$ and $E_I$. These energies were determined through fits employing Lorentzian functions. The authors interpreted the higher-energy peak ($E_I$) as analogous to an impurity band and suggested that it might represent the energy required for a transition from a bound state to the continuum. 

To analyze the possibility that $E_I$ could correspond to the energy of PF traps ($\phi_{PF}$), we extracted the tabulated value from this publication ($\delta = 0.9$, $E_I$ = 0.62 \, eV). To add additional data points corresponding to more doped samples, which have fewer oxygen vacancies, we referred to the work of Orenstein \textit{et al.}~\cite{Orenstein90}, where the optical conductivity as a function of energy is shown (see Fig. 5). Specifically, we digitized the curves corresponding to samples with superconducting transition temperatures of $T_c $= 80 \, K, 50 \, K, and 30 \, K, all of which displayed a peak in the energy region $<$ 1.25 \, eV. Since these samples are more doped than the one studied in ref[xx], they exhibit Drude-type absorption for $\omega \to 0$, which could obscure $E_J$ and possibly $E_I$ in the $T_c$ = 90 \, K sample. Figure S3 shows the digitized data along with the fitted peaks, whose values were used to estimate $E_I$. Uncertainties were associated with the peak width and were increased due to the noise introduced during digitization.  

\begin{figure}[htbp]
	\vspace{0mm}
	\centerline{\includegraphics[angle=0,scale=0.55]{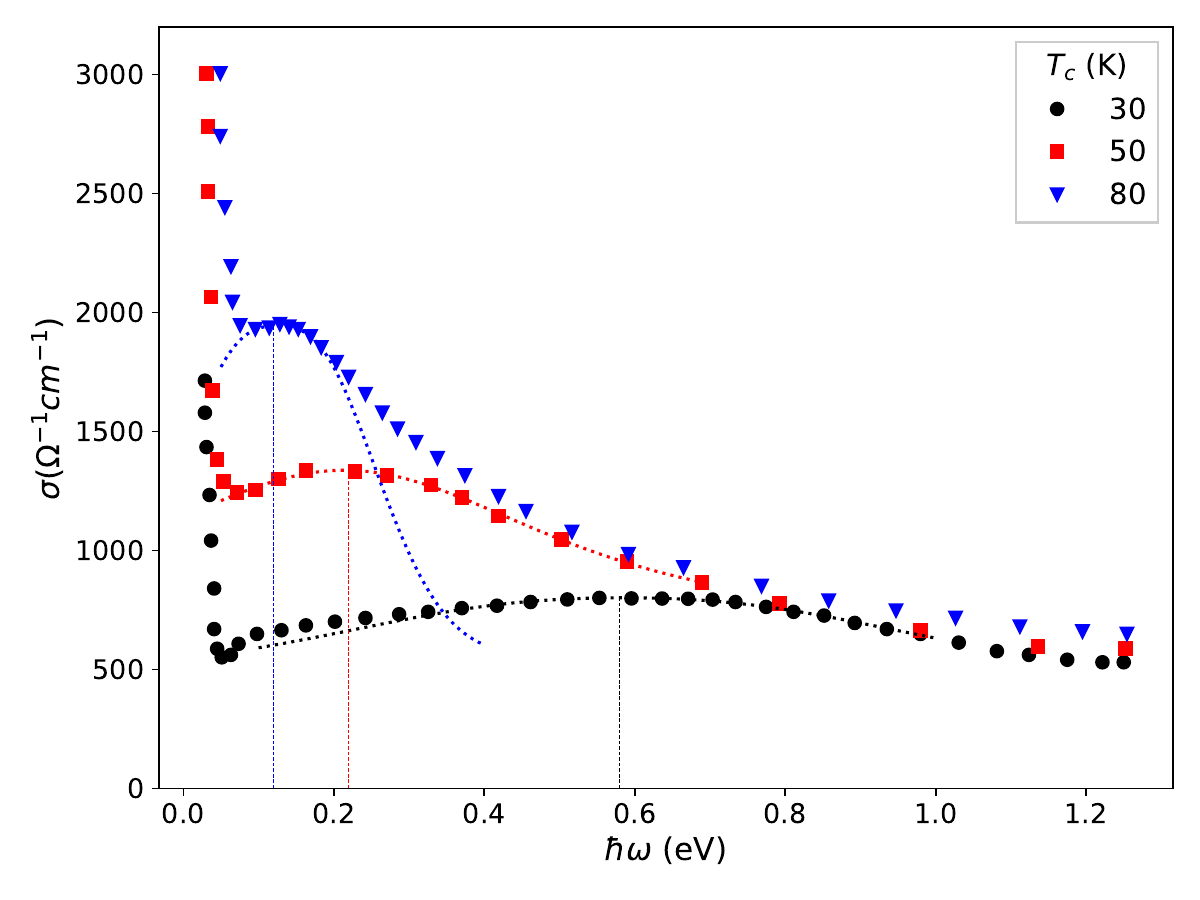}}
	\vspace{0mm}\caption{Optical conductivity as a function of the energy. The data points result from the partial digitization of curves corresponding to samples with 
		$T_c$=30 K, 50 K, and 80 K, as presented in Fig. 3 of the work by Orenstein et al.(\cite{Orenstein90}) The peak energies, determined through optimal fitting with a Pseudo-Voigt function to address the asymmetry of the observed peaks (dashed curves), are 0.12 eV, 0.22 eV, and 0.58 eV, respectively.}
	\vspace{0mm} \label{fig:sigma_vs_E}
\end{figure}

Additionally, since the authors provide the superconducting transition temperature ($T_c$) of the studied samples rather than their $\delta$ values, we relied on the works of Cava \textit{et al.}~\cite{Cava90} and Jorgensen \textit{et al.}~\cite{Jorgensen90}, where the $\delta$ value was correlated with the $T_c$ of YBCO$_{7-\delta}$ crystals using the same criterion to determine $T_c$. By digitizing Figures 1 (Cava \textit{et al.}) and 3 (Jorgensen \textit{et al.}), we obtained Figure S4. It illustrates how we estimated the $\delta$ values for the samples studied by Orenstein \textit{et al.}.

\begin{figure}[htbp]
	\vspace{0mm}
	\centerline{\includegraphics[angle=0,scale=0.55]{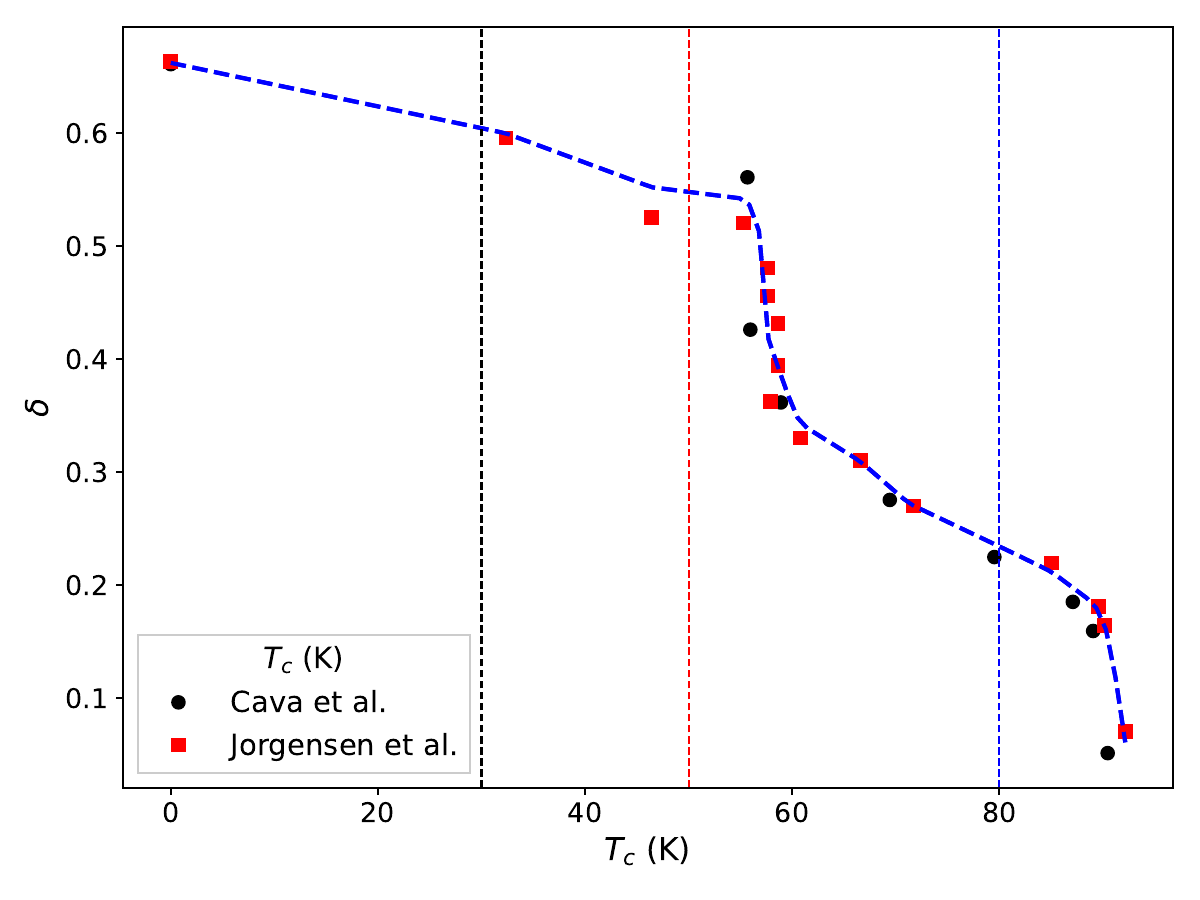}}
	\vspace{0mm}\caption{Oxygen vacancy content as a function of $T_c$ determined as the onset of magnetization measurements at low magnetic field (5 Oe). The data points were extracted by partially digitizing the original figures from Cava et al.~\cite{Cava90}  and Jorgensen et al.~\cite{Jorgensen90} . The dashed blue line correspond to the mean interpolated curve. Dashed vertical lines indicate the $T_c$ values (determined using the same method) of some samples from the study by Orenstein et al.~\cite{Orenstein90}  (30 K, 50 K, and 80 K). The average $\delta$ values obtained are 0.22, 0.55, and 0.61, respectively.}
	\vspace{0mm} \label{fig:delta_vs_Tc}
\end{figure}

\pagebreak

\end{document}